\documentclass[12pt]{article}
\usepackage[utf8]{inputenc}
\usepackage{fullpage}

\title{An overall view of key problems in algorithmic trading and recent progress}
\author{Micha\"el Karpe\thanks{Department of Industrial Engineering and Operations Research, University of California, Berkeley. Email: \href{mailto:michael.karpe@berkeley.edu}{michael.karpe@berkeley.edu}}}
\date{June 9, 2020}

\usepackage[numbers]{natbib}
\usepackage{graphicx}
\usepackage{amsmath}
\usepackage{amssymb}
\usepackage{indentfirst}
\usepackage{hyperref}

\begin{document}

\maketitle

\section*{\centering Abstract}

We summarize the fundamental issues at stake in algorithmic trading, and the progress made in this field over the last twenty years. We first present the key problems of algorithmic trading, describing the concepts of optimal execution, optimal placement, and price impact. We then discuss the most recent advances in algorithmic trading through the use of Machine Learning, discussing the use of Deep Learning, Reinforcement Learning, and Generative Adversarial Networks. \\

\section{Introduction}

Algorithmic trading is a form of automated trading with the use of electronic platforms for the entry of stock orders, letting an algorithm decide the different aspects of the order, such as the opening or closing time, the price or the volume of the order, most times without the slightest human intervention. Since algorithmic trading is used for order placement and execution, increasingly intelligent and complex algorithms are competing to optimize the placement and execution of these orders in a market that is becoming better and better understood by the developers of these algorithms. \\

\section{Key problems in algorithmic trading}

The study of algorithmic trading first requires an understanding of its fundamental issues, namely what is the best quantity for an order to be executed (optimal execution), how to place orders in a time range (optimal placement), and the consequences of these orders on the price of the stock on which we place an order (price impact). In this section, we explore these three fundamental issues using key references that address these problems. \\ 

\subsection{Optimal execution}

Optimal execution is the most known problem in algorithmic trading and was addressed by Bertsimas and Lo \cite{bertsimas1998optimal} in the case of a discrete random walk model and by Almgren and Chriss \cite{almgren2001optimal} in the case of a Brownian motion model. Optimal execution consists of buying or selling a large amount of stock in a short period, which then has an impact on the price of the stock. Such execution is optimal in the sense that one seeks to minimize the price impact or execution costs or to maximize the expectation of a predefined utility function. \\

\subsubsection{Optimal execution in Bertsimas and Lo}

In Bertsimas and Lo \cite{bertsimas1998optimal}, optimal execution is presented as an execution cost minimization problem, under the constraint of acquiring a quantity $\overline{S}$ of shares over the entire period of length $T$. Mathematically, by defining for each instant $t \in \{1, 2, \dots, T\}$, $S_t$ the number of shares acquired in period $t$ at price $P_t$, this optimal execution problem is written:
\begin{align*}
    \min_{S_t, \ t \in \{1, 2, \dots, T\}} & \mathbb{E}\left(\sum_{t = 1}^T P_tS_t\right) \\
    \text{s.t.} & \sum_{t = 1}^T S_t = \overline{S}
\end{align*}

The simplest evolution of the price $P_t$ proposed by Bertsimas and Lo \cite{bertsimas1998optimal} is to define $P_t$ as the sum of the previous price $P_{t - 1}$, a linear price impact term $\theta S_t$ ($\theta > 0$) depending on the number of shares $S_t$ acquired at the time $t$, and a white noise $\varepsilon_t$ ($\varepsilon_t \sim \text{WN}(0, \sigma^2)$):
$$P_t=P_{t-1} + \theta S_t+\varepsilon_t \ \text{ with } \  \mathbb{E}[\varepsilon_t|P_{t-1},\, S_t]=0$$

In their paper, Bertsimas and Lo \cite{bertsimas1998optimal} show that for such an evolution of the price $P_t$, the solution of the optimal execution problem is obtained recursively through dynamic programming and is $S_1^{*} = S_2^{*} = \dots =  S_T^{*} = \overline{S} / T$. \\

Then, they deal with the case of the linear price impact with information, by considering an additional term $\gamma X_t$ in the evolution of the price, such that $X_t = \rho X_{t - 1} + \eta_t$ with $\rho \in (-1, 1)$ and $\eta_t = \text{WN}(0, \sigma^2)$, as well as with the general case where $P_t = f_t(P_{t - 1}, X_t, S_t, \varepsilon_t)$ and $X_t = g_t(X_{t - 1}, \eta_t)$. In both previous cases, the solution to the optimal execution problem can still be obtained recursively through dynamic programming, resulting in a more complex formulation of the optimal execution strategy. \\

\subsubsection{Optimal execution in Almgren and Chriss}

In Almgren and Chriss \cite{almgren2001optimal}, the optimal execution problem is presented as the minimization of a utility function $U$ being the sum of the expectation $E$ and linear term of the variance $V$ of the implementation shortfall:
$$U(\mathbf{x}) = E(\mathbf{x}) + \lambda V(\mathbf{x})$$

Before defining further the expectation and the variance of the implementation shortfall, we need to define the evolution of the price and the price impact. In Almgren and Chriss \cite{almgren2001optimal}, we define $X$ the number of shares to liquidate before time $T$, $t_k = k \tau$ with $\tau = T/N$ a discretization of the time interval $[0, T]$ in $N$ intervals, $x_k$ the number of remaining shares at time $k$ and $n_k = x_{k - 1} - x_k$. The evolution of the price $P_k$ is defined as the sum of the previous price $P_{k - 1}$, a linear term of a white noise $\xi_t$ ($\xi_t \sim \text{WN}(0, \sigma^2)$) and a permanent linear price impact depending on $n_k$:
$$P_k = P_{k - 1} + \sigma \tau^{1/2} \xi_k - \tau g\left(\frac{n_k}{\tau}\right)$$
where $\sigma$ is the volatility of the asset and $g(v) = \gamma v$. \\

Almgren and Chriss \cite{almgren2001optimal} also consider a temporary price impact, however, this temporary price impact only influences the price per share $\Tilde{P_k}$ received and not on the actual price $P_k$:
$$\Tilde{P_k} = P_{k - 1} - h\left(\frac{n_k}{\tau}\right)$$
where $h(n_k / \tau) = \varepsilon \text{sgn}(n_k) + \eta n_k / \tau$, with \textit{"a reasonable estimate for $\varepsilon$ being the fixed costs of selling"} \cite{almgren2001optimal} and $\eta$ depending on the market microstructure. \\

The framework of the Almgren and Chriss \cite{almgren2001optimal} optimal execution being defined, we can now define the expectation and the variance of the implementation shortfall:
\begin{align*}
    E(\mathbf{x}) = & \sum_{k = 1}^N \tau x_k g\left(\frac{n_k}{\tau}\right) + \sum_{k = 1}^N n_k h\left(\frac{n_k}{\tau}\right) \\
    V(\mathbf{x}) = & \ \sigma^2 \sum_{k = 1}^N \tau x_k^2 \\
\end{align*}

In this framework, optimal execution strategies can also be computed explicitly and are illustrated in the form of an efficient frontier in the variance-expectation two-dimension space. Almgren and Chriss \cite{almgren2001optimal} also stress the importance of considering the risk/reward tradeoff in the calculation of optimal execution strategies, through the use of a $\lambda$ risk-aversion parameter, to create optimal execution strategies adapted to the risk profile of the executor. \\

\subsubsection{Further work on optimal execution}

Bertsimas and Lo \cite{bertsimas1998optimal} \& Almgren and Chriss \cite{almgren2001optimal} models support most of the work done on optimal execution since 2000. \\

On the one hand, recent work uses Bertsimas and Lo \cite{bertsimas1998optimal} model to show there is no significant improvement in moving from static optimal execution strategies to adapted ones for the benchmark models studied \cite{brigo2018static}. \\

On the other hand, while Almgren and Chriss \cite{almgren2001optimal} deal with the problem of optimal execution under price uncertainty, recent work uses this model to consider the problem of optimal execution under volume uncertainty, if the volume of shares that can be executed is not known in advance \cite{vaes2018optimal}. They show \textit{"risk-averse trader has benefit in delaying their trades"} \cite{vaes2018optimal} and that under both price and volume uncertainty, \textit{"the optimal strategy is a trade-off between early and late trades to balance the risk associated with both price and volume"} \cite{vaes2018optimal}. \\

\subsection{Optimal placement}

The optimal placement problem is a much less studied algorithmic trading problem than the optimal execution one. This problem consists of determining how to split the orders into the different levels of the limit order book at each period, to minimize the total expected cost. \\

This problem is summarized in Guo et al. \cite{guo2017optimal} as a problem where \textit{"one needs to buy N orders before time $T > 0$"} \cite{guo2017optimal} and where $N_{k, t}$ is the number of orders at the $k$-th best bid ($N_{0, t}$ being the number of orders at the market price), and is solved in the case of a correlated random walk model. We refer to Section 2.2. of Guo et al. \cite{guo2017optimal} for the complete formulation of the optimal placement problem. \\

\subsection{Price impact}

The price impact is mainly studied in the case of the optimal execution problem. Gatheral and Schied \cite{gatheral2013dynamical} present an overview of the main price impact models. They distinguish three distinct types of price impact: permanent, temporary, and transient. \\

\subsubsection{Permanent and temporary price impact}

Permanent and temporary price impact are usually studied together as two consequences of the same cause. Whereas the permanent price impact affects the stock price and therefore all subsequent orders, the temporary price impact only affects the price of the executed order and does not influence the stock price. Both Almgren and Chriss \cite{almgren2001optimal} and Bertsimas and Lo \cite{bertsimas1998optimal} models present permanent and temporary price impact components.

Gatheral and Schied \cite{gatheral2013dynamical} recall in their paper the notion of \textit{"price manipulation strategy"} \cite{gatheral2013dynamical}, being defined as an order execution strategy with strictly positive expected revenues. Then, they show on the one hand that an Almgren and Chriss \cite{almgren2001optimal} model which does not admit price manipulation must have a linear permanent price impact, and on the other hand that a Bertsimas and Lo \cite{bertsimas1998optimal} model with linear permanent price impact does not admit bounded price manipulation strategies. \\

An estimation of the permanent and temporary price impact in the equity market is studied in Almgren et al. \cite{almgren2005direct}, showing that while the linear permanent price impact hypothesis cannot be rejected on equity markets, the hypothesis of a square-root model for temporary impact is rejected, in favor of a power law with coefficient 0.6 \cite{almgren2005direct}. \\

However, while many articles studying the permanent price impact do so under the hypothesis of a linear impact, some research articles question this linear hypothesis, stating that permanent market impact can sometimes be nonlinear \cite{gueant2013permanent}. \\

Recent work pushes further the case of nonlinear permanent and temporary price impact, by considering a continuous-time price impact model close to the Almgren and Chriss \cite{almgren2001optimal} model but where the parameters of the price impact are stochastic \cite{barger2019optimal}. They show that their stochastic optimal liquidation problem still admits optimal strategy approximations, depending on the stochastic behavior of the price impact parameters. \\ 

\subsubsection{Transient price impact}

As explained in Gatheral and Schied \cite{gatheral2013dynamical}, \textit{"transience of price impact means that this price impact will decay over time"} \cite{gatheral2013dynamical}. Transient price impact challenges classical models of permanent and temporary price impact, especially because permanent market impact must be linear to avoid dynamic arbitrage \cite{huberman2004price}. \\

Obizhaeva and Wang \cite{obizhaeva2013optimal} propose a linear transient price impact model with exponential decay, and additional research papers also deal with the study of linear transient price impact. However, other papers show the limits of the linear hypothesis in the transient price impact model by studying the slow decay of impact in equity markets \cite{brokmann2015slow}. Recent work presents a portfolio liquidation problem of 100 NASDAQ stocks under transient price impact \cite{chen2019portfolio}. \\

\section{Recent progress in algorithmic trading}

The rise of Machine Learning over the last few years has shaken up many areas, including the field of algorithmic trading. Machine Learning, Deep Learning, and Reinforcement Learning can be used in algorithmic trading to develop intelligent algorithms, capable of learning by themselves the evolution of a stock's price or the best action to take for the execution or placement of an order. \\

In this section, we discuss the most recent advances in algorithmic trading through the use of Machine Learning, assuming that the reader already has prior knowledge in this area. First, we present applications of Deep Learning in financial engineering, then the use of Reinforcement Learning agents for optimal order execution or placement, and finally, we briefly mention applications of Generative Adversarial Networks (GANs) for financial data and time series generation. \\

\subsection{Deep Learning}

Applications of Deep Learning in financial engineering are numerous. Deep Learning is generally used for estimating or predicting financial data, such as price trends for financial products. In this subsection, we are considering a group of neural networks that are Recurrent Neural Networks (RNNs) and their applications, and the notion of transfer learning. \\

\subsubsection{Recurrent Neural Networks (RNNs)}

RNNs are a class of neural networks that allow previous outputs to be used as inputs while having hidden states. RNNs are used for sequences of data, which can correspond, for example, to a sequence of temporal or textual data. They aim to learn a sequential scheme of the data provided as the input of the network, the output of each cell depending on the output of the previous cells. \\

In financial engineering, RNNs are commonly used for stock price prediction or asset pricing \cite{chen2019deep}. Other applications include predictions of cash flows or consumer default \cite{albanesi2019predicting}, but also more original applications as part of the analysis of alternative data, with, for example, the study of textual data or the study of the evolution of satellite images to acquire information on the health of a company. \\

\subsubsection{Transfer Learning}

Transfer learning focuses on storing the knowledge gained by solving a problem and applying it to a different but related problem \cite{zhuang2019comprehensive}.  Transfer learning is generally studied in the Deep Learning framework. It aims to train a neural network on a huge dataset to have the neural network learning successfully the requested task, and then fine-tuning the training of this neural network on the few data of the new task we want our neural network to perform, this new task having generally few training data to train a model on \cite{tan2018survey}.

A very recent paper applied this concept to the transfer of systematic trading strategies \cite{koshiyama2020quantnet}. The idea proposed in this paper is to build a neural network architecture -- called QuantNet -- based on two layers specific to the market, and another layer which is market-agnostic between these two market-specific layers. Transfer learning is then carried out with the market-agnostic layer. The authors of the paper claim an improvement of the sharpe ratio of 15\% across 3103 assets in 58 equity markets across the world, in comparison with trading strategies not based on transfer learning \cite{koshiyama2020quantnet}. \\

\subsection{Reinforcement Learning (RL)}

RL is one of the three kinds of Machine Learning (along with supervised and unsupervised learning) and consists of training an agent to take actions based on an observed state and the rewards obtained by performing an action for a given state. By definition of RL, we can see algorithmic trading as an RL problem where a trading agent aims to maximize its profit from buying or selling actions taken in a market. \\

In this subsection, we first describe the challenges of using RL in algorithmic trading. We then discuss the framework of Multi-Agent Reinforcement Learning (MARL) where many trading agents compete. Finally, we explain the importance of developing a realistic simulation environment for the training of trading agents and the recent work done on this topic. \\

\subsubsection{Single-Agent Reinforcement Learning}

One of the first papers on RL for optimal execution was released in 2006 \cite{nevmyvaka2006reinforcement}, showing a significant improvement over the methods used for optimal execution, with results \textit{"based on 1.5 years of millisecond time-scale limit order data from NASDAQ"} \cite{nevmyvaka2006reinforcement}. \\

The state of an RL algorithm can have as many features as we want, however, we can easily imagine that too many features would cause the curse of dimensionality issue and features of such an RL algorithm for algorithmic trading should be chosen appropriately. Some common features used in an RL optimal execution problem are, among others, time remaining, the number of shares remaining, spread, volume imbalance, and current price. \\

The development of high-frequency trading has even made it necessary to develop RL algorithms to act quickly and optimally on the market. In recent work, the most commonly used RL algorithms for optimal execution are usually Q-Learning algorithms. Two papers published in 2018 use Q-Learning in the case of temporal-difference RL \cite{spooner2018market} and risk-sensitive RL \cite{vyetrenko2019risksensitive}. \\

Another paper released in 2018 presents the use of Double Deep Q-Learning for optimal execution \cite{ning2018double}. While Deep Q-Learning uses a neural network to approximate the Q-value function \cite{mnih2013playing}, Double Deep Q-Learning uses two neural networks to avoid overestimation that can happen when we use only one neural network \cite{hasselt2015deep}. \\

\subsubsection{Multi-Agent Reinforcement Learning (MARL)}

All previously mentioned papers are studying the optimal execution problem as a single-agent RL problem, i.e., only one agent is trained on the market data and there are no other competing agents. Such an approach is not representative of the reality of the high-frequency trading market, where not only do millions of agents train on the market and compete, but each agent is likely to adapt its strategy to the strategy of other agents. MARL is intended to address this issue by having multiple agents at the same time -- who may or may not train -- to better capture the reality of financial markets. \\

The use of MARL for market making has been addressed in a recent paper showing \textit{"the reinforcement learning agent is able to learn about its competitor's pricing policy"} \cite{ganesh2019reinforcement}. Another recent paper discusses further the need for a MARL framework for the evaluation of trading strategies \cite{balch2019evaluate}. \\

Especially, the latter reminds us that we can assess a trading strategy through two major methods, which are Market Replay and Interactive Agent-Based Simulation (IABS) \cite{balch2019evaluate}. Whereas in Market Replay, the simulation does not respond implementing the RL strategy, IABS aims to simulate responses of the market or of other agents, although an IABS simulation may remain not realistic with respect to real financial market conditions. \\

\subsubsection{On the importance of a realistic simulation environment}

After having stated the need for a MARL simulation for the training of RL trading agents, one of the key issues is to build a simulation close to the reality of the high-frequency trading market. \\

The Agent-Based Interactive Discrete Event Simulation (ABIDES) environment \cite{byrd2019abides} aims to be such a realistic financial environment, by considering the \textit{"market physics"} of the real high-frequency trading world, including a nanosecond resolution, agent computation delays or communication between agents through standardized message protocols \cite{byrd2019abides}. ABIDES also enables MARL between thousands of agents interacting through an exchange agent. \\

A recent paper studies the realism of the ABIDES environment through the use of stylized facts on limit order books \cite{vyetrenko2019real}. After a review of most of the stylized facts known for limit order books, this paper shows that the two multi-agent simulations ran into ABIDES verifies most of the tested stylized facts. However, the paper acknowledges that further improvement is needed to have all the stylized facts verified.

\subsection{Generative Adversarial Networks (GANs)}

GANs have been introduced in the Goodfellow et al. paper \cite{goodfellow2014generative}. The main idea of GANs is to train simultaneously two models, the first being called a generative model and which needs to reproduce the distribution of the data to generate, and the second being called a discriminative model and which needs to test whether a sample comes from the training data or the data created by the generative model. \\

This training process is usually presented as a two-player minimax game of a value function $V(G, D)$ which is:
$$\min_G \max_D V(D, G) = \mathbb{E}_{x \sim p_{\text{data}}(\mathbf{x})}[\log{D(\mathbf{x})}] + \mathbb{E}_{z \sim p_{z}(\mathbf{z})}[\log{1 - D(G(\mathbf{z}))}]$$

A recent paper presents the use of GANs for the generation of financial time series, with an architecture called Quant GANs \cite{wiese2020quant}. The key idea and innovation of this paper is the use of Temporal Convolutional Networks (TCNs) for the generator and the discriminator, in order to \textit{"capture long-range dependencies such as the presence of volatility clusters"} \cite{wiese2020quant}. The authors of the paper have been able to generate successfully financial time series with similar behavior than S\&P 500 stock prices.  \\

Ideally, financial data generated through GANs could be used by RL agents such as described in the previous section to improve the performance of the agents. Other applications of GANs for the generation of financial data are credit card fraud \cite{efimov2020using}, credit scoring \cite{mancisidor2020deep} or deep hedging \cite{wiese2019deep}. \\

\section{Conclusion}

Optimal execution is probably the most known problem in algorithmic trading. In this paper, we reminded the framework of the optimal execution problem in Bertsimas and Lo \cite{bertsimas1998optimal}, and in Almgren and Chriss \cite{almgren2001optimal}. We also mentioned the problem of optimal placement and discussed the distinct types of price impact, which are permanent, temporary, and transient price impact. We then described recent progress in algorithmic trading through the use of Machine Learning. Whereas the use of Deep Learning for stock prediction has already been widely explored, there is room for improvement for the use of Reinforcement Learning for algorithmic trading, and even more for the use of Generative Adversarial Networks.

\newpage

\bibliographystyle{unsrtnat}
\bibliography{main}

\end{document}